No crisis should go to waste

Mihai Nadin

Reproducibility (or lack thereof, such as in biomedical sciences, cf. Goodstein[1], or notoriously in psychology) affects more than the validation of experiments. Often, assumed reproducibility justifies transforming experimental evidence into convenient law ("evidence-based" policy, cf. Sarewitz[2]). Failed reproducibility—almost exclusively in life science experiments—stands in contradistinction to experiments in "physics or astronomy or geology" (Goodstein). These knowledge domains are identified as test provable, not by accident, as we shall see.

The crisis of reproducibility, which the American National Academies is examining as well, should not go to waste. It is an opportunity for opening a discussion of the relation between various knowledge domains and the need to adapt research methods to the specific dynamics of the subject that scientists attempt to describe.

Not yet articulated—to the best of my knowledge—is the call to the scientific community to re-evaluate the underlying assumptions upon whose basis knowledge acquisition and confirmation are pursued. Massively failed reproducibility has encouraged finger-pointing and palliatives, but not the critical re-evaluation of the epistemological perspective. In particular domains, 80% of published results, from researchers who earned the respect of their peers, proved to be irreproducible. Therefore the thought that something might be off with the expectation that research, no matter which subject or purpose, is best validated through reproducible experiments cannot be wished away. The understanding of what Newton called *Nature*, under which label he aggregated both the physical and the living, might prove as inadequate in our time as it was when it was articulated.

After vitalism was debunked, science rejected the distinction between the living and the non-living. This in itself is quite surprising, since in science you don't throw away a question because it was improperly answered. The foundational works in defining the living of Walter Elsasser[3] and Robert Rosen[4] (not to mention Schrödinger[5]), advancing views of nature different from those of Newton and his followers, were pretty much ignored at the time they were published. Their arguments, quite different in their perspectives, deserve a closer look at this moment of questioning research and validation methods of life sciences. The living is heterogenous, purposeful, and anticipatory; the non-living is homogenous, purpose-free, and reactive. If indeed, to know is to be aware of distinctions—especially those of fundamental nature—variations cannot be eliminated by fiat.

While physics and physics-based disciplines (such as chemistry) adequately describe the non-living, there remains a need for a complementary perspective that expresses the nature of life. What defines this perspective is that the specific causality characteristic of life is accounted for by integrating past, present, and possible future. The living changes in a way different from the non-living.

Taking Gödel's concept of decidability (the logic pertinent to axiomatic systems used in arithmetic operations) and applying it to defining knowledge domains is an opportunity.

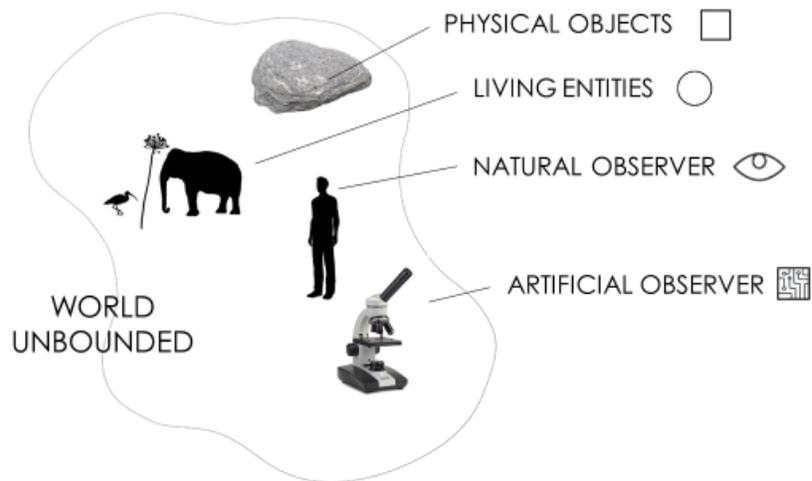

Fig. 1. Unbounded world as open system

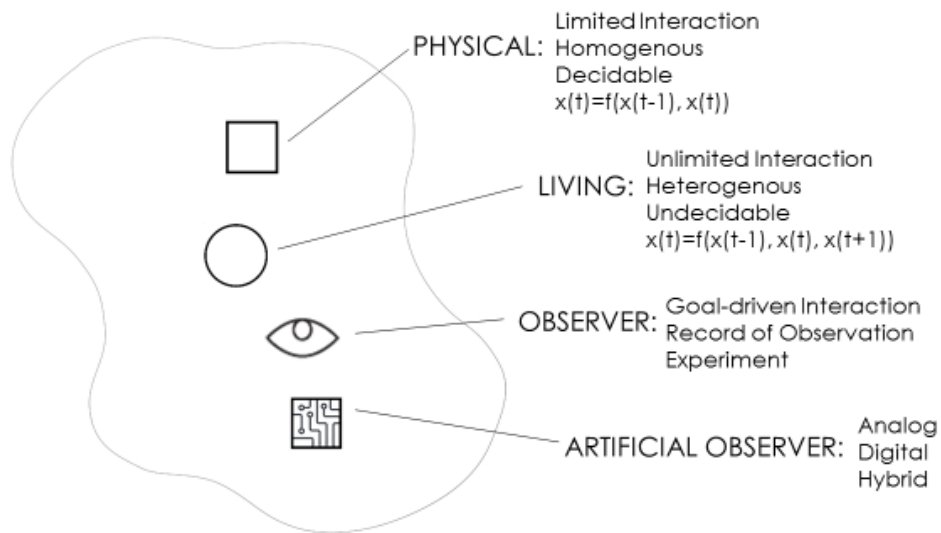

Fig. 2. Characterizing the physical and the living

But the focus in this alternative view is not on Gödel's rigorous logical proof, as it is on the notion of decidability, extended here from the formal domain to that of reality.

Definition: A subject is decidable if it can be fully and consistently described. Indeed, physics, astronomy, geology (mentioned by Goodstein), knowledge domains where reproducibility is close to 100%, represent descriptions of dynamics (how things change) that can be complete and consistent. Such descriptions undergird predictions—the expected output of science.

Thesis 1: The threshold from the decidable to the undecidable is the so-called *G-complexity* (G for Gödel, obviously; Nadin[6]).

Thesis 2: Change is the outcome of interaction.

The living, in its unlimited variety of ever-changing forms is G-complex, i.e., characterized by undecidability. For non-living physical entities, interaction takes the specific form of deterministic reaction, expressed in physical laws (such as those expressed in Newton's equations or in Einstein's theory of relativity). For the living, change is the outcome of interactions in which the physical (the dynamics of action-reaction) is complemented by anticipatory expression: current state contingent upon possible future state. Living interaction is not reducible to the physical action-reaction sequence. The description of physical interaction conjures quantity, and results in data. The description of living interaction conjures quality, and results in information, i.e., data associated with meaning (Wheeler[7]).

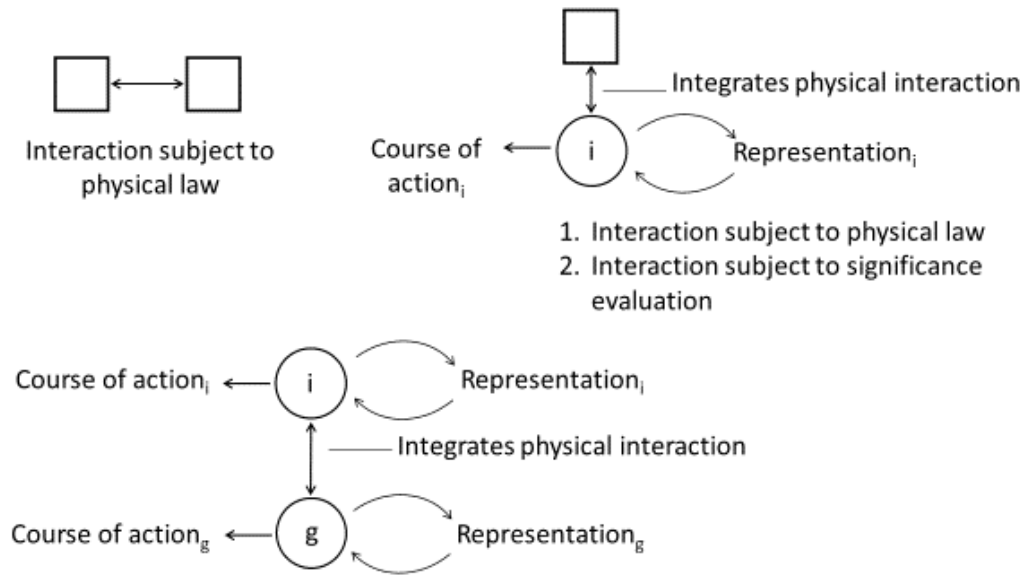

Fig. 3. Change is the outcome of interaction

As a consequence, to expect experiments involving the living (of interest not only to psychology, but also to the biomedical sciences and many other fields of inquiry pertinent to life) to be reproducible is epistemologically equivalent to reducing the living to its physical substratum, and biology to physics and chemistry. Information, characteristic of life, is not physical (Lopez-Suarez[8]). Too often, such experiments turn out to be mere instances of conditioning (psychology outperforms every other known discipline in this respect). The outcome is more testimony to the ignored limits of perception (the well-documented time resolution of one tenth of a second, cf. Canales[9]) and to how well the subject was conditioned. This limited understanding of causality is occasionally transcended in modern science (and not only in the quantum mechanics perspective). Nevertheless, it is still the dominant view. (Just take a look at the Call for Proposals in the biological sciences that the National Science Foundation puts out.)

Mapping from an open system (extending from the cell to the whole human being) of extreme dynamics to the closed system of the experiment (which by definition is supposed to be decidable) might result in reproducibility. But what is reproduced is a false assumption, not a path to knowledge about change. The validity of some 40,000 fMRI studies, and more broadly the interpretation of neuroimaging results, was recently questioned (Eklund, Nichols, Knutsson[10]), after the fMRI (25 years old) technology itself was critically assessed (Shifferman[11]). False-positive rates of up to 70% concerning its most common statistical methods, which have not been validated using real data, are actually a proof of a misguided assumption.

More recently, brain activity has become the object of computational modeling. There is much to gain from computational models in physics applications (the Juno space mission is only a recent example). Their intrinsic limitation in studying living processes stems from the fact that algorithmic computation captures only the deterministic aspects of change. Therefore the guaranteed reproducibility of computational neuroscience experiments conjures knowledge and validation not about the brain, whose deterministic and non-deterministic aspects complement each other, but about algorithmic computation. Interactive computation, in line with the dynamics of interaction of the living in general, and of the brain in particular, is rarely considered (Nadin[12, 13]).

As impressive as the Human Genome project was, it is a good example of irreproducible experiments. It was generated under the reductionist assumptions of a blueprint—published as such (*Science*, 16 February 200, Vol. 291, Issue 5507)—of a *homo sapiens* that does not change over time, i.e., epigenetics ignored. What was extracted is a truncated image of gene syntax. The 1000 Genomes Project (2008-2015), aimed at studying variation (initially ignored) and genotype data, is an example of an improved understanding but yet another irreproducible experiment. It affords empirical data, i.e., access to some semantic aspects of gene expression. The goal, probably not yet on the radar of scientific inquiry, should be the pragmatic level, where meaning is constituted in the context of life unfolding in an anticipatory manner.

This idea is relatively well illustrated by the entire cycle of reproduction. Pregnancy (Brunton, Russell[14]) is a convincing example of anticipatory expression underlying creation, i.e., the birth of some entity that never existed before. Actually, the living is in a continuous state of remaking itself, *sui generis* re-creation of its constitutive cells—each different from the other—and thus of the entire organism. The constancy of physical (non-living) entities, even those of extreme dynamics (such as black holes), stands in contrast to the variability of any and all organisms and the matter in which they are embodied.

An assumption similar to that of the Human Genome governs the current Connectome project. It will be ten or one hundred times more costly than the Genome project, but not better in reporting on the variability of the cortex. Windelband's[15] view of nomothetic science—expressed in universally valid laws (such as Newton's laws of mechanics)—and idiographic science—diachronic processes subject to empirical observations—could as well guide in defining new methods for gaining knowledge peculiar to the living. Bernstein[16] wrote about the "repetition without repetition" characteristic of the living as an expression of its dynamic variability. This is yet another argument in favor of finally transcending the machine view characteristic of Cartesian determinism and reductionism. Gelfand's[17] take on the matter points in the same direction: "There is only one thing which is more unreasonable than the unreasonable effectiveness of mathematics in physics, and this is the unreasonable ineffectiveness of mathematics in biology." Progress in science renders the need for a "new Cartesian revolution," at the forefront of science's efforts to better understand change in the specific manner in which it characterizes life.

References

1. Goodstein, D. Scientific Misconduct, *Academe*, January-February, 28-31 (2002)

2. Sarewitz, D. Reproducibility will not cure what ails science (World View), *Nature* 525:7568, 159 (2015)

3. Elsasser, W. *Reflections on a Theory of Organisms. Holism in Biology*. Baltimore: Johns Hopkins University Press, 1998

4. Rosen, R. *Life Itself. A Comprehensive Inquiry into the Nature, Origin, and Fabrication of Life* (Complexity in Ecological Systems). New York: Columbia University Press, 1991

5. Schrödinger, E. *What is Life?* New York: Macmillan, 1944

6. Nadin, Mihai. G-Complexity, Quantum Computation and Anticipatory Processes, *Computer Communication & Collaboration*, 2:1, 16-34. Toronto: BAPress (2014)

7. Wheeler, J.A. Information, Physics, Quantum: the Search for Links, Proc. 3rd Int. Symp. Foundations of Quantum Mechanics, 354-368. Tokyo, 1989

8. López-Suárez, M., Neri, I., Gammaitoni, L. Sub-$k_BT$ micro-electromechanical irreversible logic gate, *Nature Communications* 7, 28 June 2016

9. Canales, J. *A Tenth of a Second: A History*. Chicago: The University of Chicago Press, 2009

10. Eklund, A., Nichols, T.E., Knutsson, H. Cluster failure: Why fMRI inferences for spatial extent have inflated false-positive rates, *PNAS*, 113:28, 7900-7905

11. Shifferman, E. More Than Meets the fMRI: The Unethical Apotheosis of Neuroimages, *Journal of Cognition and Neuroethics* 3 (2): 57-116

12. Nadin, M. Anticipation and the Brain. In: Nadin, M. (ed.) *Anticipation and Medicine*. Cham: Springer (forthcoming 2016)

13. Nadin, M. Anticipatory and Predictive Computation. In: Laplante, P. (ed.) *Encyclopedia of Computer Science and Technology*. London: Taylor and Francis, 2016

14. Brunton, P.J., Russell, J.A. The expectant brain: adapting for motherhood, *Nature Reviews Neuroscience* 9, 11-25

15. Windelband, W. *Geschichte und Naturwissenschaft. Rectoratsreden der Universität Strassburg* [History and Natural Science, Rectoral Address], 1894, 355-379. Tübingen: Mohr, 1907

16. Bernstein, N.A. The coordination and regulation of movements. Oxford: Pergamon Press, 1967. (See also: Nadin, M. (ed) *Learning from the Past. Early Soviet/Russian contributions to a science of anticipation*. Cognitive Systems Monographs. Cham CH: Springer, 2015)

17. Gelfand, I.M. In: Borovik, A.V., *Mathematics Under the Microscope. Notes on Cognitive Aspects of Mathematical Practice*. September 5, 2007. Creative Commons, http://eprints.ma.man.ac.uk/844/ [accessed 20 October 2015]